\documentclass[12pt]{iopart}

\usepackage{epsfig}
\usepackage{graphicx}
\usepackage{iopams}
\usepackage{color}

\begin{document}

\title{Competition and fragmentation: a simple model generating
lognormal-like distributions}
\author{V. Schw\"ammle$^1$, S. M. D. Queir\'{o}s$^{1,2}$, E. Brigatti$^{1,3}$ and T. Tchumatchenko$^4$}
\address{$^1$Centro Brasileiro de Pesquisas F\'isicas, Rua Dr. Xavier Sigaud 150,
22290-180, Rio de Janeiro, RJ, Brazil.}
\address{$^2$Unilever R\&D Port Sunlight, Quarry Road East, CH63 3JW, Wirral, UK.}
\address{$^3$Instituto de F\'{\i}sica, Universidade Federal Fluminense, Campus da Praia
Vermelha, 24210-340, Niter\'oi, RJ, Brazil.}
\address{$^4$Institute for Applied Physics, TU Darmstadt, Schlossgartenstr. 7, D-64289 Darmstadt, Germany.}

\begin{abstract}
The current distribution of language size in terms of speaker population is
generally described using a lognormal distribution. Analyzing the original
real data we show how the double-Pareto lognormal distribution can give an
alternative fit that indicates the existence of a power law tail. A simple Monte Carlo model is constructed
based on the processes of competition and fragmentation. The results reproduce the power law tails of the real
distribution well and give better results for a poorly connected topology of interactions.
\end{abstract}
\maketitle

The astonishing similarity between biological and language evolution has
attracted the interest of researchers familiar with analyze genetic
properties in biological populations with the aim of describing problems of
linguistics \cite{Cavalli97,Cavalli01,Maynard95}. Their techniques, for
instance, have succeeded in explaining some interesting features related to
the coexistence of the about $7,000$ languages present on Earth.

More recently, the effort to connect evolutionary biology with linguistics has emerged into a study that describes the effects of
competition between languages on language evolution. The work of Abrams and
Strogatz in 2003~\cite{Abrams2003}, who analyzed the stability of a system
composed of two competing languages, can be considered as the starting point
of this new research line. In the following years, other groups,
simultaneously, developed new analytical and computational models~\cite
{Patriarca2004,Mira2005,Kosmidis2005,Schulze2005b,Stauffer2005,Schwaemmle2005b,Pinasco2006,deOliveira2005}.
An overview of the fast increasing literature on language competition can
be found in refs.~\cite{Stauffer2006,Schulze2008}.

Languages are by no way static. They continuously evolve, changing, for
example, their lexicon, phonetic, and grammatical structure. This evolution
is similar to the evolution of species driven by mutations and natural selection~\cite{Sereno91}.
Following the common
picture of biology \cite{Spagnolo2004}, changes in the language structure may be seen as the
result of microscopic stochastic
changes caused by mutations.
Natural selection, which may be caused by competition between
individuals,
positively selects some of these small changes, depending on their reproductive success.
A sequence of macroscopic observations corresponds to such a microscopic picture.
In language evolution, these macroscopic events are, for instance,
the origination of two languages from an ancestor
one - for example, the emergence of the Romance languages from Latin - or
the extinction of a language.

\begin{figure}[tb]
\centering\includegraphics[angle=0,width=0.8\textwidth]{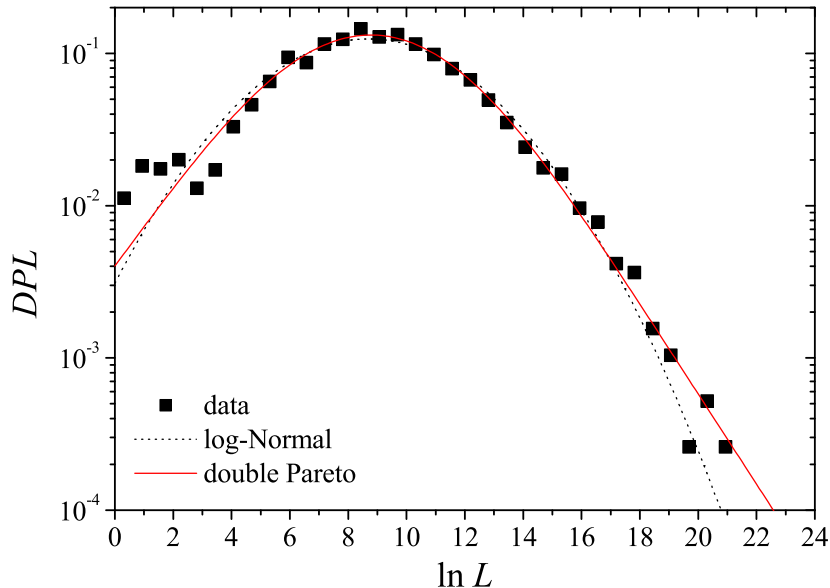}
\caption{DPL distribution and respective
fits with a double-Pareto lognormal distribution (continuum line) and lognormal one (dotted
line). We counted the number of languages with speaker populations between 
$\exp(0.625 n)$ and $\exp(0.625(n+1))$ for
$n=0,1,2,3,...$. Although the double-Pareto lognormal distribution shows the same deviation from the real data for small
values of $L$, it presents a better adjustment of the whole curve when in
comparison to the log-normal fit (see $D_{KS}$ values in the text). In particular,
it is possible to see how the fat tail for languages with a big population size
is well fitted by a power law. 
}
\label{fig:real}
\end{figure}

In this work we model the evolution of languages from a macroscopic
point of view. More precisely,
the microscopic processes responsible for the differentiation of one language into two new languages are not implemented here.
Effectively, we neglect the microdynamics that
generates language changes, at the level of individuals, and we just
describe their effect on extinction and differentiation at the level of
languages, throughout a phenomenological mechanism of growth and fragmentation.
Language change is determined by the dynamics of the size of its population.
The fact that rare languages
are less attractive for people to both learn and use is the mechanism
considered as the origin of these population size changes. Consequently, this
mechanism introduces a sort of frequency dependent reproductive success for different
languages. Statistical data supporting this conjecture can be found in Ref.~%
\cite{Sutherland2003}. Languages documented as declining are negatively
correlated with population size. This phenomenon is similar
to the Allee effect in biology~\cite{Stephens99}.
With a simple computational model, based on the above described mechanisms, we compare
simulation results with empirical data of the distribution of population sizes of languages (DPL) of Earth's actually spoken
languages~\cite{Grimes2000}  (see Fig.~\ref{fig:real}).

Several attempts have recently been made to reproduce the DPL.
Two works focused on the apparently lognormal shape of the DPL.
Tuncay~\cite{Tuncay2008} described language differentiation by means of a
process of successive fragmentations, in combination with a multiplicative
growth process. In a recent paper by Zanette~\cite{Zanette2007}, the
dynamics of language evolution is considered as a direct consequence of
the demographic increase of the speaker populations, which is modeled by
means of a simple multiplicative process. Unsurprisingly, these models
obtain pure lognormal distributions for the DPL, as expected from the
application of the central limit theorem for multiplied random variables \cite{Mitzenmacher2004a,Mitzenmacher2004b}.
Unfortunately, the DPL is known to significantly differ from a pure lognormal shape~%
\cite{Sutherland2003}.

The Schulze
model~\cite{Schulze2005b} relies on ideas already successfully applied to
model biological evolution. Languages are identified by a bit string that
represents their characteristic features. New languages are produced by mutations of these features and small
languages are discriminated by competition. These
simulations, during the transient towards the stationary state, are able to
generate data with a distribution similar to the DPL.
A review of the
Schulze model and its application to different problems connected to
language interaction can be found in Ref.~\cite{Schulze2008}.

Another model, the Viviane model, simulates human settlement on an unoccupied region.
Languages suffer local mutations, until the available space becomes
completely populated~\cite{deOliveira2005}. The introduction of a bit string
representation into the Viviane
model was able to generate new results which reproduce
the DPL over almost the entire range well, except for large language sizes~\cite%
{Oliveira2007,Oliveira2008}. The bit string approach gives an explanation for the deviation of the DPL from a lognormal distribution for small population sizes. The DPL changes its shape depending on the method used to distinguish different languages from dialects. When restricting the comparison between languages to be based on a small number of different features in language structure, the deviation for small language populations appears. This interpretation was confirmed by using a simple model~\cite{Schwaemmle2009} that neglects the geographic effects present in the Viviane model.

A thorough look at the DPL may suggest that the deviations from the lognormal
shape could be due to power law decays~\cite{Gomes99}.
We investigate this idea
by fitting the DPL with a double-Pareto lognormal distribution and additionally comparing
it to the simulation results of our new model.

The paper is organized as follows. First we analyze the DPL and show that
the double-Pareto lognormal distribution \cite{Reed2002,Reed2004} gives
an alternative fit. The next section
introduces the computational model. In the last two sections, the simulation results are presented and
the conclusions are given, respectively.

\section{Statistical analysis of the distribution of languages}

Reference~\cite{Grimes2000} provides
the number of people speaking a given language as their own
mother tongue. There, 6,912 different languages have been classified and for 6,142  of them their
speaker population was estimated. In the data set of 6,142 languages we also verify
a big difference between the median, $6000$ speakers, and the average of the number speakers of a language,
circa $1.052 \times 10^6$ speakers. This discrepancy is related to the fact that only 326 languages have at
least one million speakers. When they are assembled, these languages account for more than $95 \% $ of
world's population with the remaining 5816 ($94.6 \%$) languages encompassing the left-over speaking
population. This draws attention to the fat tailed behaviour of the distribution of the number of
speakers. 
Starting from these data, we make a histogram by counting the number of languages with
population size enclosed in a bin with values between $\exp(0.625 n)$ and $\exp(0.626(n+1))$
where $n=0,1,2,...$.
This distribution defines the DPL.
A pure lognormal shape can be described by,
\begin{equation}
DPL_{\textrm{lognormal}}(L) = A \exp \left[ -\frac{(\ln L-\mu )^{2}}{2\,\sigma ^{2}}\right]
\label{eq:DPL}
\end{equation}
where $A$, $\mu $ and $\sigma $ are the parameters of the distribution.
A lognormal distribution corresponds to a parabola in a double--logarithmic plot (see
Fig.~\ref{fig:real}).

In order to calculate the parameter values by maximum log-likelihood estimation
we have to transform the
distribution to obtain a probability density function. As the size of the bins in the DPL increases
exponentially, the frequency of languages with a certain population size is calculated by dividing the DPL by the bin width which leads to,
\begin{equation}
c(L)=\frac{1}{\sqrt{2\,\pi \,\sigma ^{2}}\,L}\exp \left[ -\frac{(\ln L-\mu
)^{2}}{2\,\sigma ^{2}}\right] ~,  \label{eq:lognormal}
\end{equation}%
after further normalization in order to satisfy $\sum_L c(L) = 1$.

The fit by maximum log-likelihood estimation with the lognormal distribution of eq.~(\ref{eq:lognormal}) gave the
parameter values $\mu =8.70$ and $\sigma =3.20$ with a
Kolmogorov-Smirnov distance of $D_{KS}=0.0253$. These values of the lognormal distribution are the same as the ones obtained in ref.~\cite{Sutherland2003}, where $\log_{10}(L)$ was used instead of $\ln(L)$.

Unfortunately, as stated before, the DPL of real data is known to significantly differ from
lognormality \cite{Sutherland2003}
and deviations can be easily observed for small and large values of $L$.

In particular, a thorough look on the DPL shows that it does not seem to
decay exponentially. Hence, we have tried
to fit the data using a distribution that can account for such a power law
decay. As both tails could show power laws, we assumed the
so-called double Pareto lognormal distribution, introduced by Reeds~\cite%
{Reed2002,Reed2004},%
\begin{equation}
c\left( L\right) =\frac{\alpha \,\beta }{\alpha +\beta }\left( A\,L^{-\alpha
-1}\Phi \left[ \frac{\ln L-\nu -\alpha \,\tau ^{2}}{\tau }\right]
+B\,L^{\beta -1}\Phi ^{\prime }\left[ \frac{\ln L-\nu +\beta \,\tau ^{2}}{%
\tau }\right] \right) ,  \label{doublepareto}
\end{equation}%
where $\Phi \left( x\right) $ represents the cumulative distribution
function of the normal distribution up to $x$ ($\Phi ^{\prime }\equiv 1-\Phi
\left( x\right) $), $A=\exp \left[ \alpha \nu +\alpha ^{2}\frac{\tau ^{2}}{2}%
\right] $, and $B=\exp \left[ -\beta \nu +\beta ^{2}\frac{\tau ^{2}}{2}%
\right] $.

In this case, using the maximum log-likelihood procedure, we have
estimated the following values for the function parameters: $\alpha =0.682$,
$\beta =0.603$, $\tau =2.33$, and $\nu =8.89$. This set of parameters yields
$D_{KS}=0.0198$ which is slightly smaller than the distance presented by the
log-normal fit. With these values we can compute the critical $P$ value,
$P_{c}$, which is given by
\begin{equation}
\sqrt{\frac{-\ln \left( \frac{P_{c}}{2}\right) }{2\,N}}=D_{KS}.
\label{eq:Pc}
\end{equation}
The closer $P_{c}$ is to zero, the poorer the fit is. With the values
presented previously we have $P_{C}\simeq 1.6\times 10^{-2}$ for the fit for
Eq. (\ref{doublepareto}) and $P_{C}\simeq 8\times 10^{-4}$ for the
log-normal adjustment.
This fact suggests that the double Pareto distribution allows a better
description of real data. Figure~\ref{fig:real} shows the results of this
analysis comparing the corresponding DPLs.

It is likely to obtain better results by using a fitting function with a larger number of adjustable parameters. The Akaike information criterion~\cite{Akaike74} provides an estimation whether the introduction of new parameters into the fitting procedure is useful,
\begin{equation}
AIC = 2 k +  N \log{\frac{RSS}{N}}~.
\end{equation}
$k$ denotes the number of fitting parameters and $RSS$ is the residual sum of squares. Smaller values of this criterion correspond to better results. We obtain $AIC=-10.25$ for the fitting with the lognormal function and $AIC=-12.08$ for the one with the double-Pareto lognormal function. 


\section{Interaction versus fragmentation}

\noindent
This section introduces our model. As we stated before, we describe the
behavior of languages on the macroscopic scale of sub-populations,
neglecting the languages internal structure, constituted by the speaking
preferences of each individual.
Each language $i$ is characterized exclusively by the number of its speakers $L_{i}$.
The origination of one new language is obtained throughout fragmentation.
Fragmentation is implemented
as follows: at each time step, each language can break
into two new languages with a fixed probability $F$:
\begin{equation}
 L_i(t) \longrightarrow L_i(t+1)=\frac{1}{2}
 L_i(t)\,,\,L_{new}(t+1)=\frac{1}{2} L_i(t) \,.
\end{equation}
For the sake of simplicity, the two new languages contain
exactly half of the population of the ancestor language (fragmentations into
two parts of unequal size do not alter the qualitative shape of the
distribution of the fragments size). Taking into account only the
fragmentation process, the number of languages, $N_{L}$, increases until
each language has only one speaker. For large numbers of $N_{L}$, there is an
interval, during the time evolution, when the population distribution
displays a pure lognormal shape.

It is interesting to remember that a simple soluble model analog to this fragmentation
process is the discrete sequential fragmentation of a segment. In this case,
through a rate equation approach, it is possible to show that an explicit
asymptotic solution is given by a lognormal distribution. This is
independent on the number of pieces of each breaking event~\cite{Delannay96}.

The interaction between languages is implemented by a term that controls the
growth of each language in dependence of its relative population size. At each time step,
the population of each language $i$ follows the rule:
\begin{equation}
L_i(t+1) = 
\cases{
L_i(t) + \frac{I}{N_{Tot}(t)} \sum \limits_{j\textrm{\tiny : NN or all}} (L_i(t) - L_j(t))
- \frac{N_{Tot}(t)}{V} & for $L_i(t+1) \geq 1$ \cr
\textrm{language removed} & otherwise\cr
}
 \label{eq:1}
\end{equation}
where $L_i(t)$ denotes the number of speakers of language $i$ at time step $t$, $I$ the
interaction strength and $N_{Tot}(t) = \sum \limits_{i=0}^{N_L(t)} L_i(t)$ is the
total number of speakers. The second term, which can be positive or negative,
enhances the reproductive success of the languages with more speakers, and
causes the decrease of the number of rare language speakers. The third term,
which is controlled by the parameter $V$, is a cause of random death that avoids the
unlimited growth of the population. The term limits the growth of each language to have less than $V$ speakers. It is important to notice
  that this third term is necessary because the second one causes an
  uncontrolled growth in the total population. In fact, the second
  term does not conserve the total population because languages with
  less than one speaker ($L_i(t)<1$) are removed from the
  simulation. 

We perform the simulations over two different topologies. In the first, the
languages are ordered in a chain, with periodic boundary conditions, and
interact with their nearest neighbors (NN).
New languages, generated by fragmentation, are placed between the
ancestor language and the following one. This implementation describes
restricted local interactions, that can only happen between neighbor
languages, in a 1-dimensional space.

In the second implementation, the interaction is carried out between all
languages, in a mean field like model. For each iteration,
every single language interacts with the whole set of languages. This model
corresponds to the ideal case of a fully connected world, without geographic
constraints.

\begin{figure}[tb]
\centering
\includegraphics[angle=270,width=0.6\textwidth]{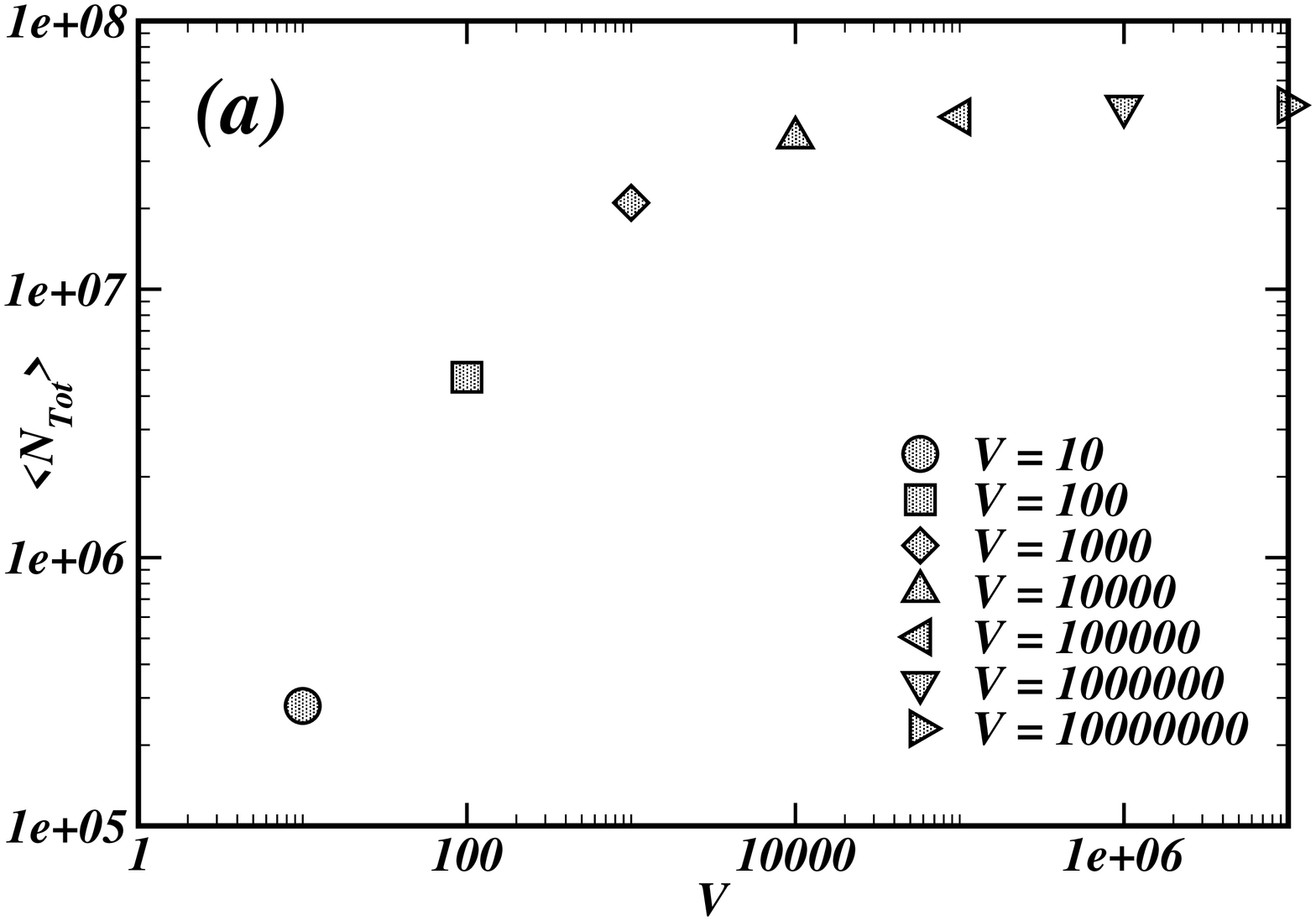} %
\includegraphics[angle=270,width=0.6\textwidth]{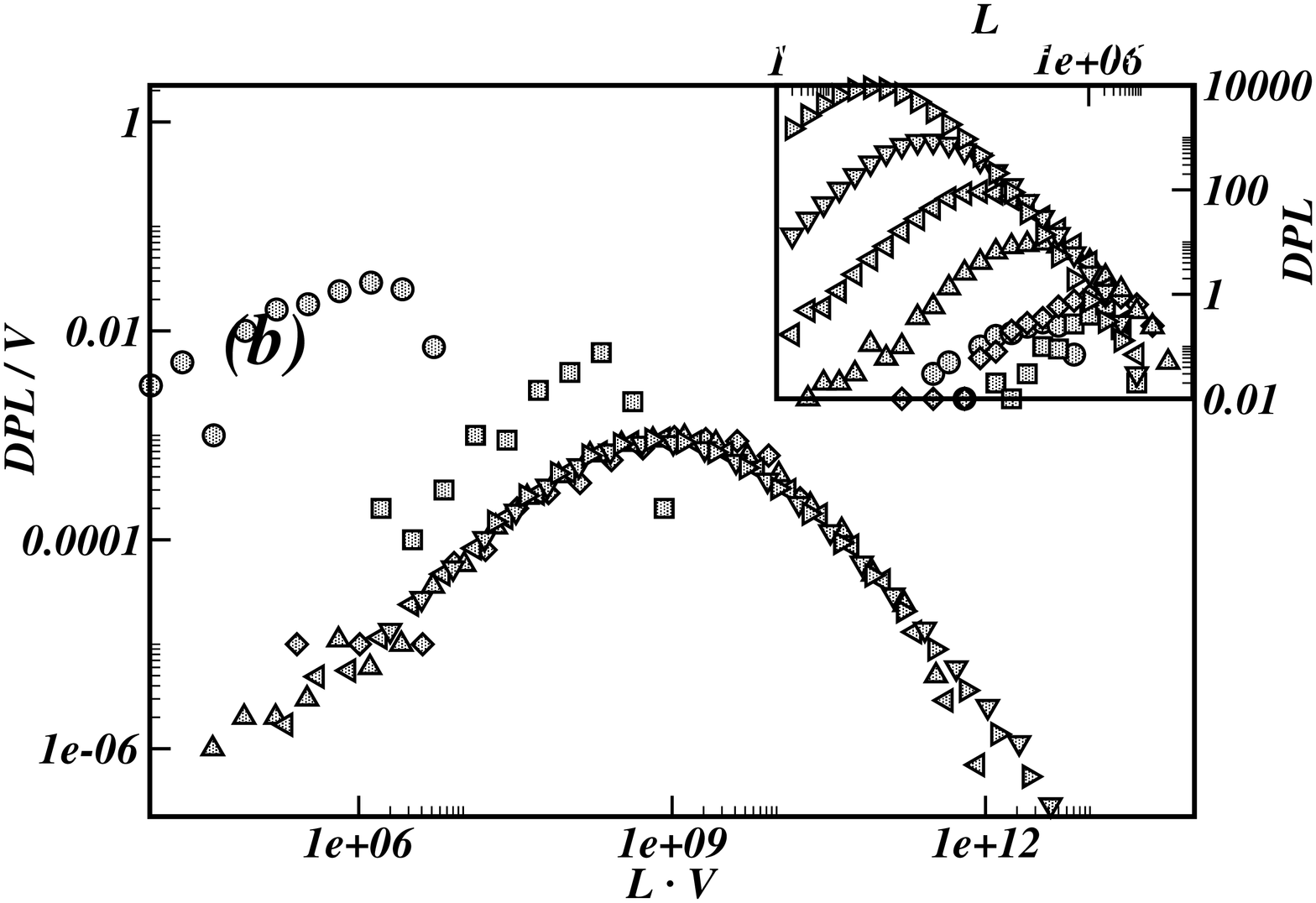}
\caption{Local interactions. {\bf (a)}: Averaged total population $<N_{Tot}>$ for
different values of $V$ and fixed $I=1,000,000$ and $F=0.1$. For large
$V$ (we took the values $V=10,100,1000,...,10^7$), the
total population saturates. {\bf (b)}: Rescaled population distribution for
different values of $V$ (same symbols and same parameter values as in {\bf (a)}.
The distributions can be collapsed for large values of $V$. Inset: The
same but unscaled population distributions.}
\label{fig:Chain_V}
\end{figure}

\section{Simulation results}

\noindent All our simulations runs begin with one language having a number
of speakers that can vary from $1,000$ to $1,000,000$. However, the final
state of the system does not depend on the initial conditions.
The system is characterized by three global quantities: the number of
languages $N_L$, the total population $N_{Tot}$ and the distribution of
language sizes in terms of speaker population. This distribution is averaged
over time steps. The results
are collected after the stationary state is reached, i.e. with both $N_L$ and $N_{Tot}$
fluctuating around some fixed value. Note, that eq.~(\ref{eq:1}) allows the case $N_{Tot}>V$ for $N_L\geq2$.\\

\noindent

\paragraph{Local interactions\newline}

\noindent We start with languages arranged on a 1--dimensional array. This
setup can be interpreted as the most simple implementation to describe small range
interactions corresponding to a weakly connected world.

\begin{figure}[tb]
\label{fig:Chain_NTot_NL}\centering
\includegraphics[angle=270,width=0.6%
\textwidth]{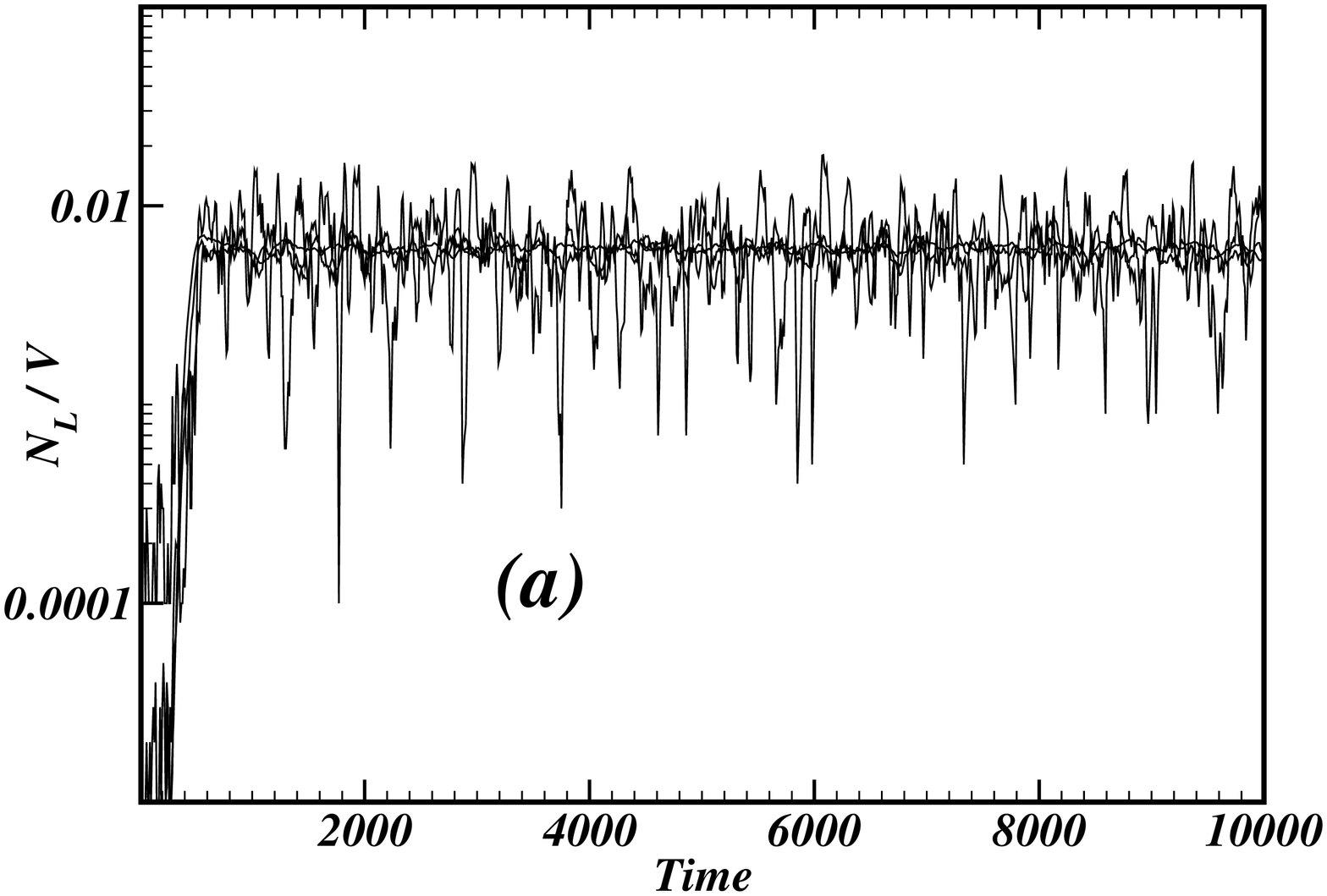} %
\includegraphics[angle=270,width=0.6\textwidth]{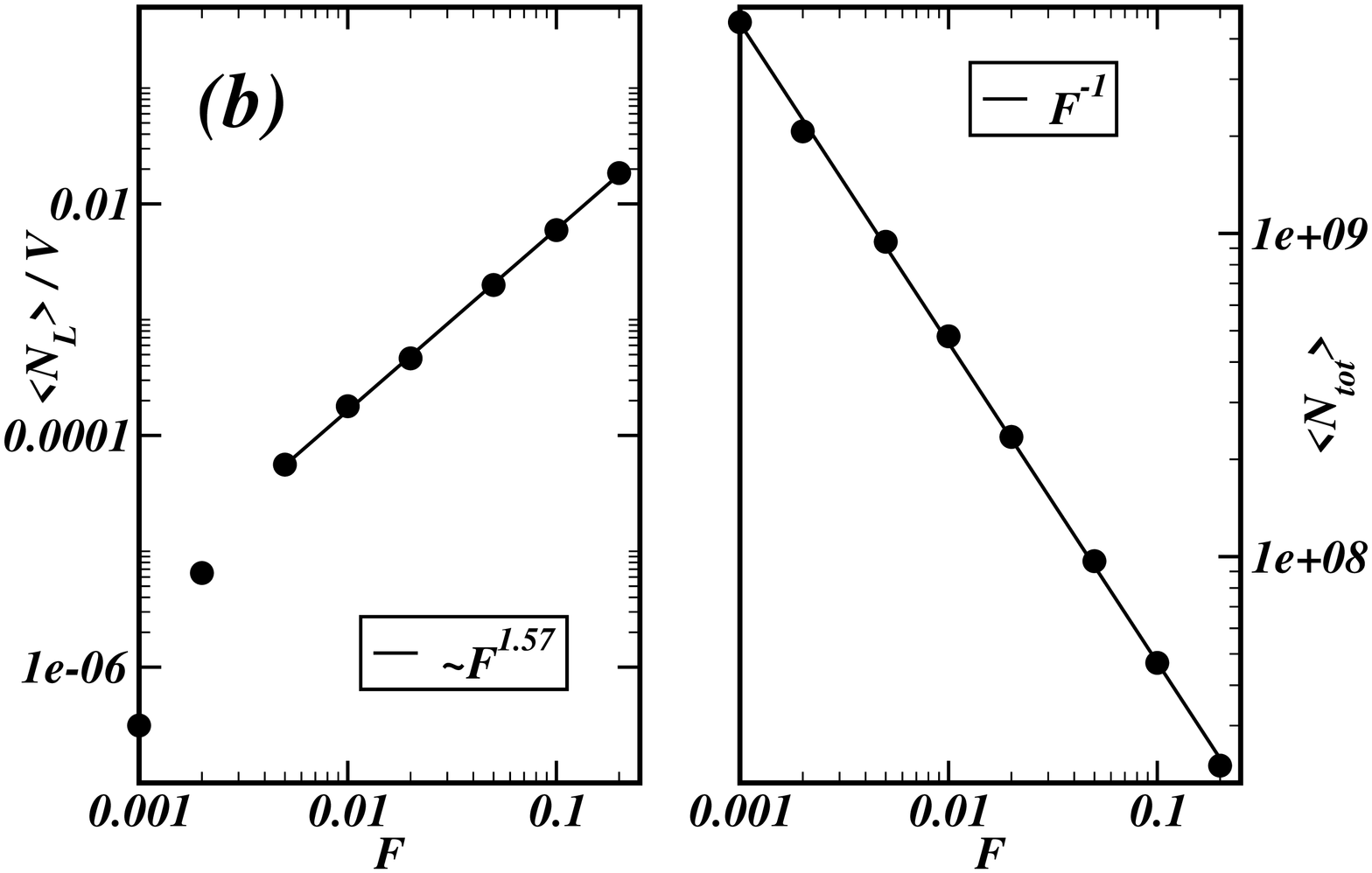}
\caption{Local interactions. {\bf (a)}: the normalized number of languages $N_L/V$
plotted versus simulation time steps. These data indicate that $N_{L}$
increases linearly with $V$.
Parameter values are $I=10^{6}$, $F=0.1$ and different $V$ values between $10000$ and $10^7$.
{\bf (b)}: Averaged total population size $<N_{Tot}>$ and the normalized
averaged number of languages $<N_L>/V$ versus the fragmentation probability $%
F$ ($I=10^6$). }
\end{figure}

First, we study the model dependence on the parameter $V$. Without the
interaction term, this parameter would directly determine the population size,
and it is normally called carrying capacity or Verhulst factor in models of population dynamics.
The presence of the second term in Eq.~(\ref{eq:1}) generates a
different behavior. As it can be observed in Fig.~\ref{fig:Chain_V}a, the
averaged total population size saturates for large $V$ values. In contrast, after a short transient,
for sufficiently large values of $V$, the number of languages, $N_{L}$,
increases linearly with $V$, as shown in Fig.~\ref{fig:Chain_NTot_NL}a.

Looking at Fig.~\ref{fig:Chain_V}b, we can see how, for large $V$ values,
it is possible to rescale all the DPL towards a  common scaling behavior.
The $V$ dependence of the distributions becomes clear when looking at the inset of the same figure.
In fact,  for small $V$ one language completely dominates the system.
By increasing $V$, the interaction term becomes weaker, and the
fragmentation process takes over, leading to a larger
number of languages. For even larger $V$ values, the distribution moves
towards smaller languages sizes maintaining its shape when displayed in a double-logarithmic scale as shown in
Fig.~\ref{fig:Chain_V}b.

In Fig.~\ref{fig:Chain_NTot_NL}b, we show the power law dependence of
the total population size and of the number of languages with the parameter
$F$, which controls the fragmentation process.

\begin{figure}[tb]
\centering
\includegraphics[angle=270,width=0.6\textwidth]{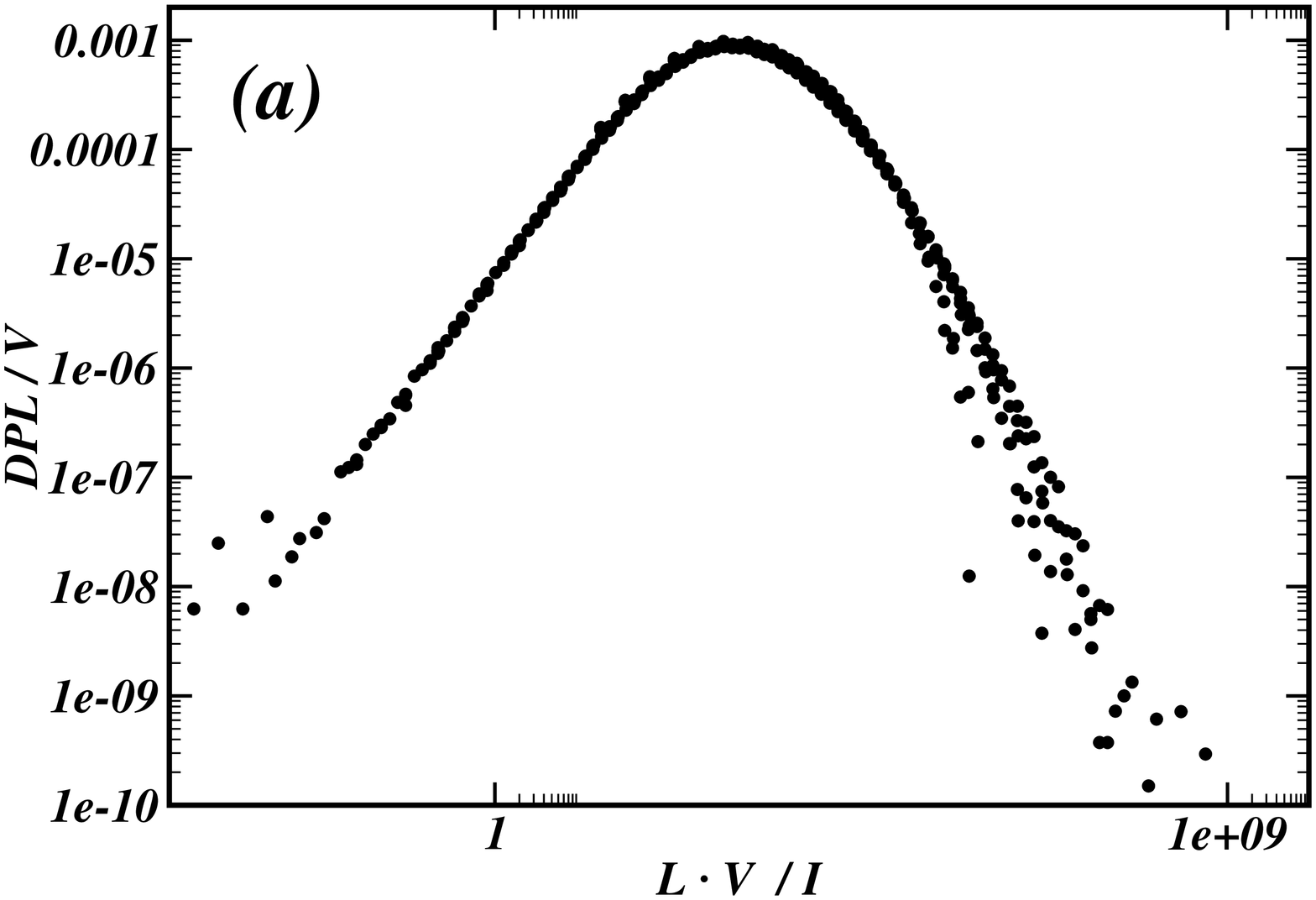} %
\includegraphics[angle=270,width=0.6\textwidth]{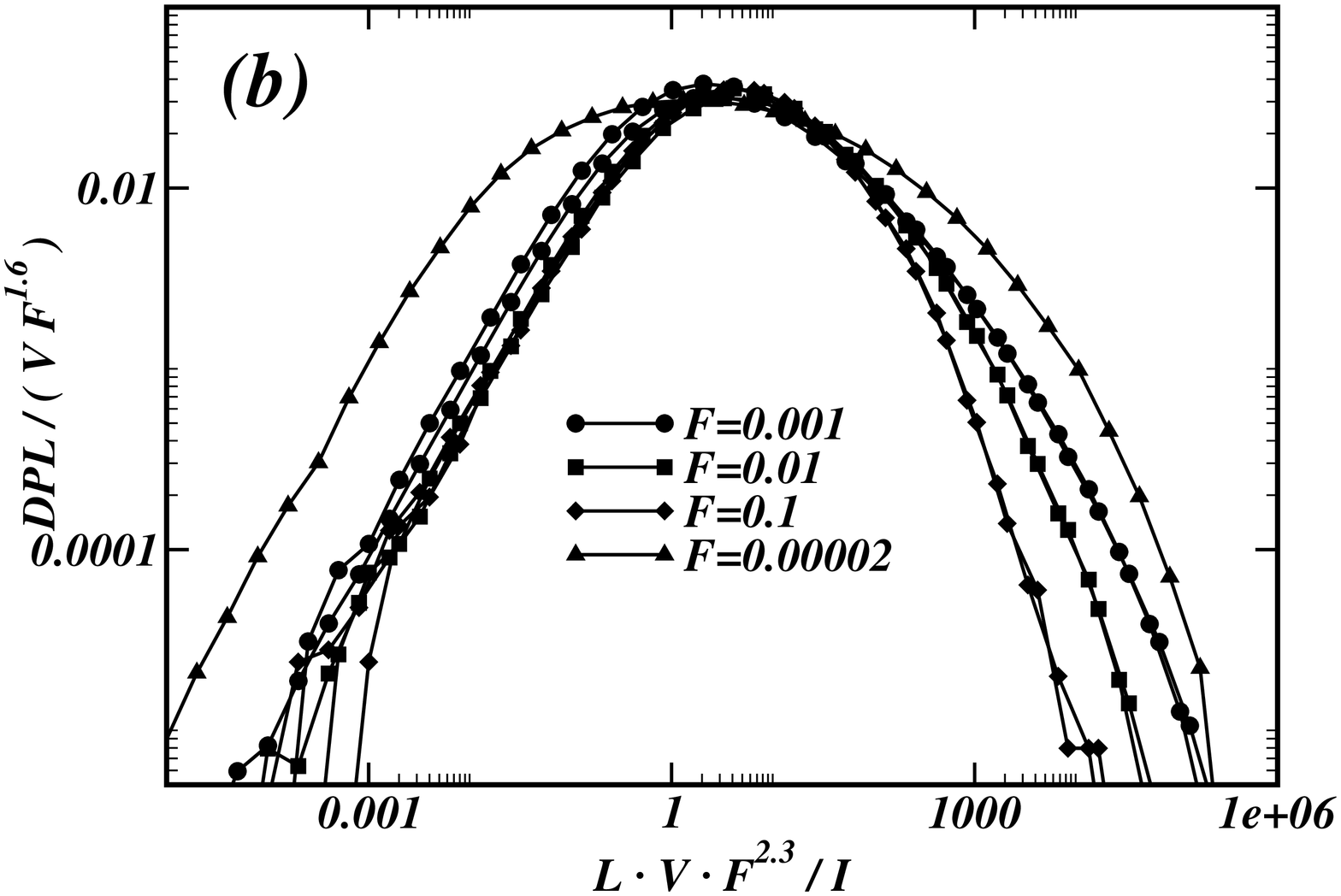}
\caption{Local interactions. {\bf (a)}: Histogram collapse for different $I$ and $%
V$. Apparently, the shape is not lognormal but a combination of two power
laws. The parameter ranges for the 17 simulations are
$I=10^{3}-10^{8}$ and $V=10^4-10^7$. $%
F $ is fixed and equals $0.1$. {\bf (b)}: Rescaled histogram for different
values of the fragmentation probability $F$. The F-scaling in
  the horizontal axis is introduced in order to collapse the position of the
  maxima for different simulations. The parameter ranges are $%
I=10^{3}-10^{7}$, different values of $V$ and $F=2\cdot 10^{-5}-0.1$. The
smaller $F$, the broader is the distribution.}
\label{fig:Chain_I}
\end{figure}

Figure~\ref{fig:Chain_I}a shows the shape of the DPL for different values of the
parameter $V$ and of the interaction strength $I$. It is possible to obtain
a data collapse by rescaling the distribution with the value of $V$ and the
population size with the factor $V/I$.
It is interesting to note that, in log-scale, the broadness of the
distribution does not depend on $V$ or $I$.
As can be clearly observed, the shape of the curve deviates strongly from a
lognormal shape for small and large population values. In fact, we collected data
over sufficiently many decades to be able to show that the curves decay like
a power law for small and large population sizes.

Figure~\ref{fig:Chain_I}b  shows the shape  of the distribution for
different values of all parameters of the model. The fragmentation
probability $F$ changes the width of the distributions. For smaller $F$ we
obtain broader distributions. From the analysis of this data we obtain
scaling relations for the maximum value of the distribution (scaling with
$VF^{2.3}/I$) and for its normalization ($VF^{1.6}$).  This last scaling relation
follows exactly from  the $V$ and $F$ dependence on the averaged number of languages, $N_L$, shown in Fig.~\ref{fig:Chain_NTot_NL}b.

\begin{figure}[tb]
\centering
\includegraphics[angle=270,width=0.8\textwidth]{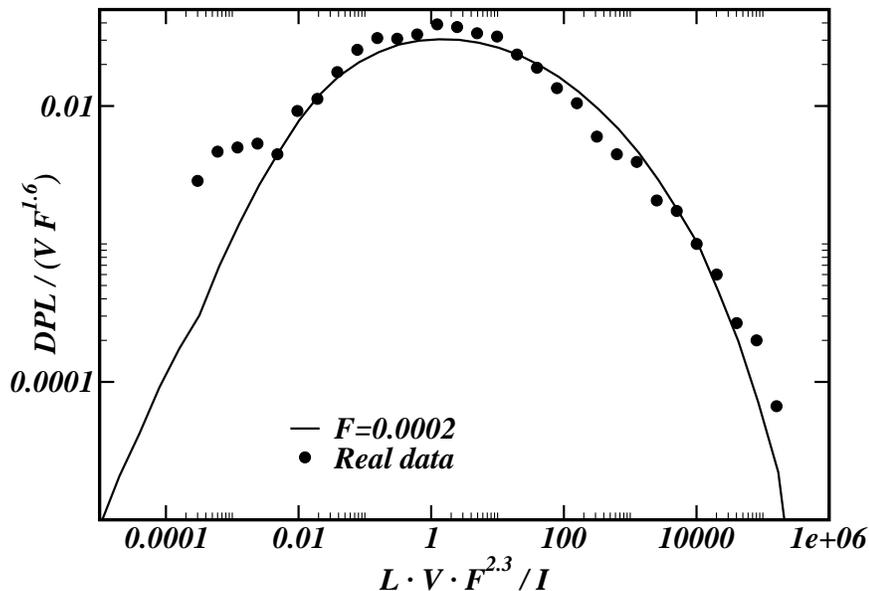}
\caption{Local interactions. The real data (circles) is rescaled with the
values $I=1.5 \cdot 10^3$, $V=5 \cdot 10^{11}$ and $F=2 \cdot 10^{-5}$.
These parameters correspond to the one that allow rescaling of real data with the ones shown in Fig~\ref{fig:Chain_I}.
The continuous line represents the rescaled simulation results.
The parameters used in
the simulation are: $I=10^7$, $V=3.2\cdot 10^{12}$ and $F=2 \cdot 10^{-5}$.}
\label{fig:CompReal}
\end{figure}

Using these scaling relations we can estimate the parameters values for
running a simulation that can reproduce the DPL of real data. In fact, these
parameters correspond to the ones that allow the rescaling of the DPL of real data to collapse with the ones shown
in Fig~\ref{fig:Chain_I}b.
Taking $F=2\cdot 10^{-5}$ from adjusting the broadness of the distribution, the values of the others
parameters are $I=1.5\cdot
10^{3}$, $V=5\cdot 10^{11}$.
 Unfortunately, these values correspond to
simulations with too demanding computational time.
For this reason, we are forced to just test the quality of the collapse of
rescaled real data and the rescaled simulation, and we can not do a direct
comparison. We carry out a simulation with the parameters $F=2\cdot
10^{-5},I=10^{7}$ and $V=3.2\cdot 10^{12}$.
In Fig.~\ref{fig:CompReal}, we can see how the DPL collapses well with
our simulation if we neglect very small population sizes ($L<100$). In
addition, our model can reproduce the behavior of the
fat tail well (which corresponds to frequently spoken languages).\\

\noindent
\\

\paragraph{Mean field\newline}

\noindent

The increasing connection between people speaking different mother languages
suggests exploring the behaviour of our model located on a topology with more links than
those of a 1--dimensional array. For this reason, we decided to explore the other
limiting situation: a fully connected model (mean field like description).
Even if this is a quite unrealistic implementation, we would like to estimate the upper bound
for our dynamics, corresponding to a fully globalized world.

\begin{figure}[tb]
\centering
\includegraphics[angle=270,width=0.6\textwidth]{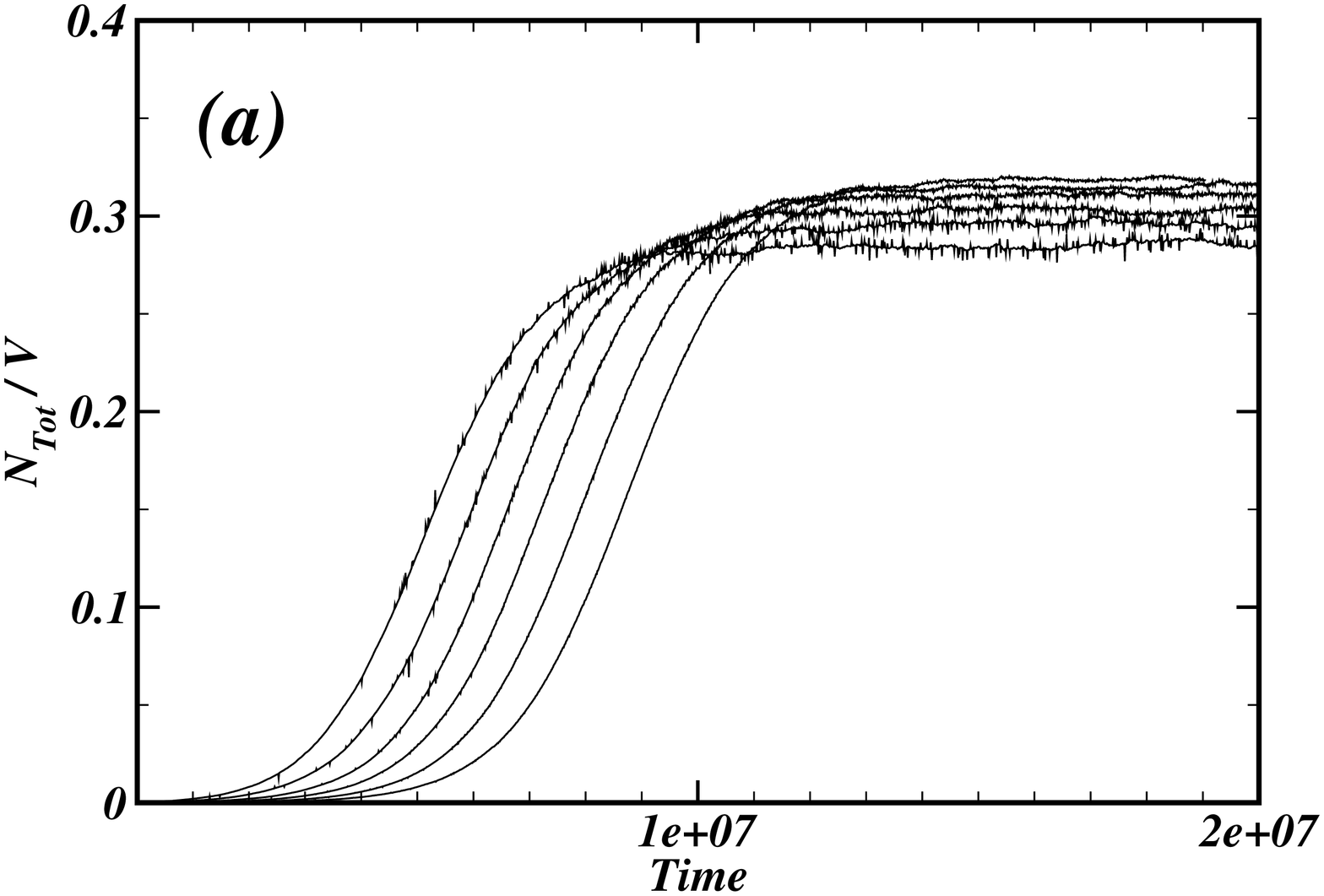} %
\includegraphics[angle=270,width=0.6\textwidth]{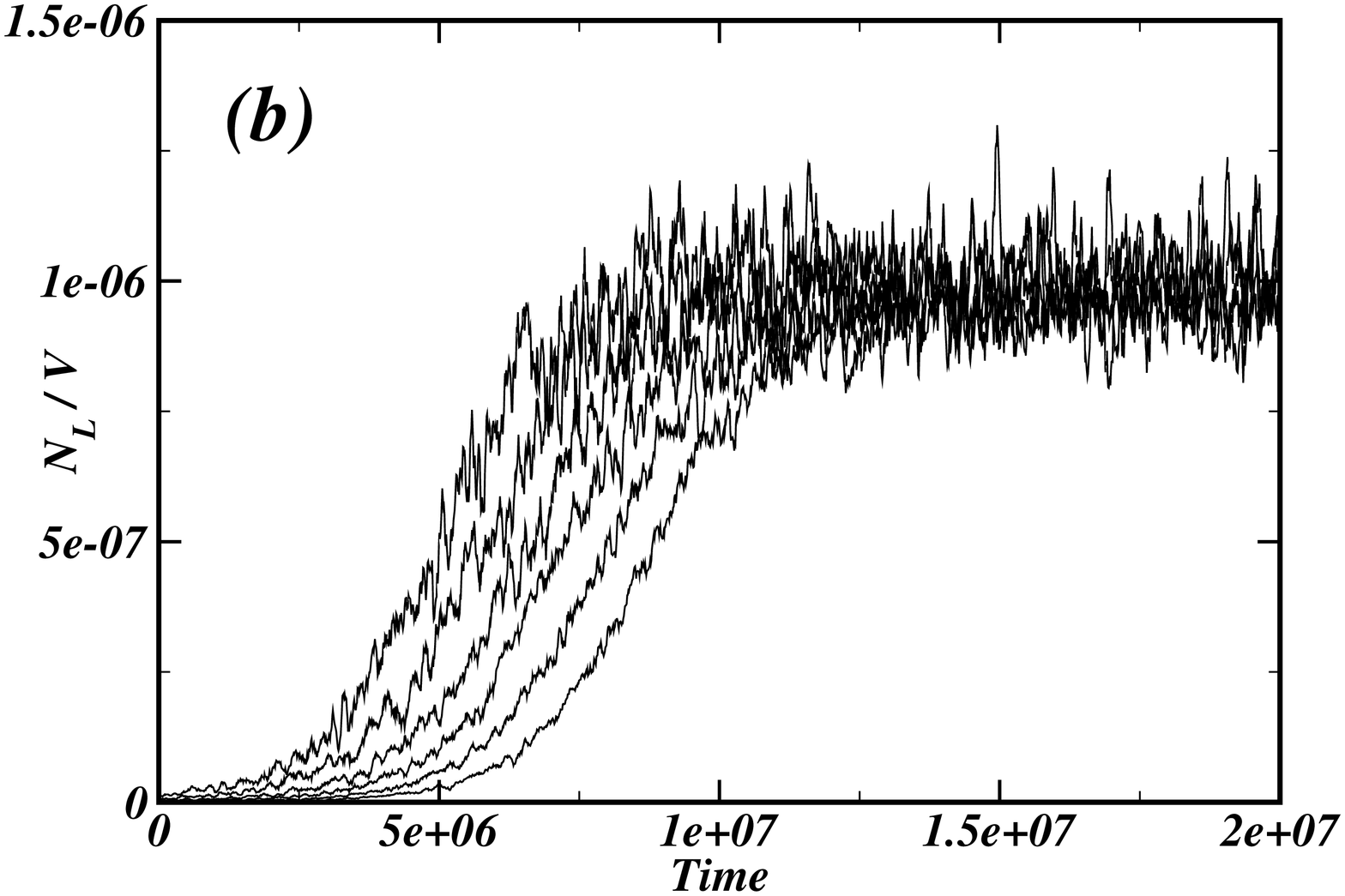}
\caption{Mean field interactions. The normalized total population {\bf (a)} and
the normalized number of languages {\bf (b)} versus simulation time steps.
We fixed $I=100$ and
$F=0.001$ and varied $V$ from $10^8$ to $32\cdot 10^8$. The figures show
that both quantities approximately increase linearly with $V$. For smaller $V$, the
stationary state is reached faster.}
\label{fig:Means}
\end{figure}

Figure~\ref{fig:Means} shows the normalized total population, $N_{Tot}/V$,
and the normalized language number, $N_L/V$, for different $V$ values. In
this topology, both quantities approximately grow linearly with $V$ and $N_{Tot}$ does
not reach any saturation value. This novel behavior is due to the
different form of carrying out the sum in the second term of Eq.~\ref{eq:1}. In fact, in this implementation,
the sum is taken over all languages and not only over the two neighbors, making the sum to have values of the
order of $N_{tot}$.

\begin{figure}[tb]
\caption{Mean field interactions. Data collapse for the distribution of
language size in terms of speaker population. Sufficiently large total
populations are considered (large $V$ values). Also for this implementation,
the distributions show fat tails, with exponents 1 and -2. The parameter
ranges are within $I=10-100,000$, $V=10,000-10^9$ and $F=0.001-0.1$. }
\label{fig:MFVerh}\centering
\includegraphics[angle=270,width=0.8\textwidth]{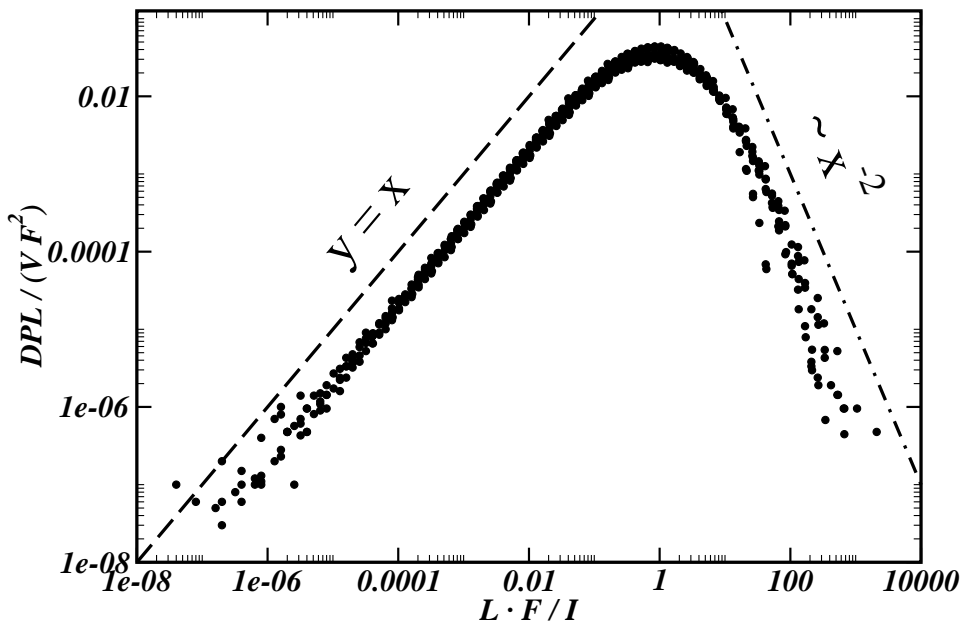}
\end{figure}
For large values of
$V$, the distribution approaches a fixed lognormal-like shape. In Fig~\ref{fig:MFVerh}, we show the data collapse for different values
of the parameters $V$, $I$ and $F$. Surprisingly, the scaling
relations are really simple and all DPLs can be perfectly collapsed
into each other. On the logarithmic scale, the width of the distribution remains
fixed and it does not change when varying the parameters values. Even the
$F$-dependence, leading to a different broadness of the DPL in the previous implementation,
disappears. The
distribution clearly shows two power law decays with exponents $1$ and $-2$.

These results are quite intriguing. On the one hand, the fixed value of the
distribution broadness does not allow the reproduction of the real data
of actual language sizes. On the other hand, it suggests an interesting
conjecture. If the increasing connection between people, in the future, 
will allow the description of their interaction using a mean field like
approximation, our model may provide some predictions. Independently on the
population growth, the width of the DPL distribution will get narrower.
This means that, despite of the probable growth of the world population,
more languages will become extinct than new languages appear. The
number of less spoken languages will strongly decrease.

\section{Conclusions}

A careful analysis of the distribution of population sizes of languages (DPL)
suggests that the behavior of its tail is
better described by a power law. For this reason, a fit
of the real data using the so-called double-Pareto lognormal distribution
generates better results than the fit with a pure lognormal distribution.

These deviations from a pure lognormal distribution mean that simple models
that reproduce exactly such distributions
must be improved. We implemented
a new toy model on a 1-dimensional topology that, starting from a
macroscopic description at the sub-population level, accounts for two
general mechanisms: language competition and fragmentation. These
ingredients are sufficient to generate distributions that well reproduce
real data, particularly the behavior of frequently spoken languages
(the right tail of the distribution).

Moreover, we studied our model on a fully connected network, in which all
sub-populations are in mutual contact (mean field like behavior).
This implementation can be a good approximation for
describing the trend for the interaction patterns on a future,
characterized by more and more interconnected communities. This
model version allows a simple prediction that confirms previous
conjectures about the future mass extinction of languages~\cite{Sutherland2003}: despite of the
probable growth of the world's population, more languages will become
extinct than new languages will appear.

Finally, we want to recall that, as our toy model is based on simple and general rules,
  it may be characteristic for systems other than language evolution as well.
In fact, systems with an interplay between a fragmentation process and a size dependent growth should
exhibit similar  patterns.
A good example comes from the description of the growth of companies \cite{Stanley96,Stanley98a,Stanley98b}
or from the analysis of the mutual fund size distribution \cite{Farmer08}.
In these systems it is possible to describe the size distributions in terms of
log-normal-like distributions 
In fact, as a zeroth order approximation, log-normal distributions can naturally be generated
by a multiplicative growth processes, in which
the company size, at any given time, is given as a multiplicative factor times the size of the company at a previous time.
For the logarithm of the company size, this process becomes an additive process
and so the distribution converges to a normal distribution, obeying the central limit theorem (Gibrat's law
\cite{Gibrat} and theory of breakage~\cite{Kolmogorov41}).
Small variations in this approach, as the introduction of some type of cutoff, can modify the lognormal
distribution into a heavy tailed distribution \cite{Mitzenmacher2004a}. 
This is obtained, for example, by introducing creation and annihilation processes \cite{Simon58,Gabaix03}.
At this level of description, none of the components of the random process depend on size.
Anyway, it is well known that the
growth rates of companies are size dependent \cite{Stanley96,Stanley95, Bottazzi2003}
and once size effects are taken into
account the predictions for the
distribution can become more quantitative \cite{Farmer08}.
This description of the growth of companies is similar to our approach,
in the sense that the dynamics
of creation and annihilation in  growing phenomena
can be analogously described introducing
the process of fragmentation
as well as its counterpart, the phenomenon of coalescence.
Our work points out the importance of taking into account these mechanisms,
in modelling systems usually described in terms of birth and death processes
or random growth \cite{Saichev2008}.


\section*{Acknowledgements}
We thank G. Savill for critical reading. V.S. and E.B. benefited from financial support from the Brazilian
agency CNPq and S.M.D.Q. from European Union through BRIDGET Project (MKTD-CD 2005029961). 

\section*{References}
\bibliographystyle{unsrt}
\bibliography{language}


\end{document}